\documentclass[sigconf,10pt]{acmart}

\RequirePackage[utf8]{inputenc}

\usepackage{booktabs} 

\usepackage{enumitem}

\usepackage{hyperref}           
\hypersetup{
    colorlinks=true,
    linkcolor=blue,
    filecolor=red,      
    urlcolor=magenta,
    breaklinks=true,            
}

\definecolor{dark-orange}{HTML}{A01C00}
\definecolor{light-orange}{HTML}{FFA592}
\definecolor{light-blue}{HTML}{74DDFF} 
\definecolor{dark-blue}{HTML}{00698B}

\usepackage{color}
\newcommand\fixme[1]{}

\copyrightyear{2018} 
\acmYear{2018} 
\setcopyright{rightsretained} 
\acmConference[CrossCloud'18]{5th Workshop on CrossCloud Infrastructures \& Platforms }{April 23--26, 2018}{Porto, Portugal}
\acmBooktitle{CrossCloud'18: 5th Workshop on CrossCloud Infrastructures \& Platforms , April 23--26, 2018, Porto, Portugal}\acmDOI{10.1145/3195870.3195873}
\acmISBN{978-1-4503-5653-4/18/04}

\urldef\acmurldoi\url{https://doi.org/10.1145/3195870.3195873}

\renewcommand{\thefootnote}{\fnsymbol{footnote}}

\newcommand\blfootnote[1]{%
  \begingroup
  \renewcommand\thefootnote{}\footnote{#1}%
  \addtocounter{footnote}{-1}%
  \endgroup
}

\begin{document}

\setlength{\pdfpageheight}{\paperheight}
\setlength{\pdfpagewidth}{\paperwidth}

\title{Fair non-monetary scheduling in federated clouds}

\author{Miłosz Pacholczyk}
\affiliation{%
  \institution{Institute of Informatics, University of Warsaw}
  \city{Warsaw}
  \country{Poland}
}
\email{mwpacholczyk@gmail.com}

\author{Krzysztof Rzadca}
\affiliation{%
  \institution{Institute of Informatics, University of Warsaw}
  \city{Warsaw}
  \country{Poland}
}
\email{krz@mimuw.edu.pl}

\begin{abstract}
In a hybrid cloud, individual cloud service providers (CSPs) often
have incentive to use each other's resources to off-load peak loads or
place load closer to the end user. However, CSPs have to keep track of
contributions and gains in order to disincentivize long-term
free-riding. 
We show CloudShare, a distributed version of a
load balancing algorithm DirectCloud based on the Shapley value---a powerful fairness
concept from game theory.
CloudShare coordinates CSPs by a ZooKeeper-based coordination
layer; each CSP runs a broker that interacts with local
resources (such as Kubernetes-managed clusters). 
We quantitatively evaluate our implementation by simulation. The
results confirm that CloudShare generates on the average more fair
schedules than the popular FairShare algorithm. 
We believe our results show an viable alternative to monetary methods
based on, e.g., spot markets.
\end{abstract}

%
%

\ccsdesc[500]{Computer systems organization~Cloud computing}
\ccsdesc[300]{Software and its engineering~Process management}
\ccsdesc[300]{Computing methodologies~Planning and scheduling}

\keywords{scheduling, distributed, fairness, Shapley value,
  cooperative game theory, ZooKeeper, Kubernetes}

\maketitle

\section{Introduction}

In a hybrid cloud (supercloud)~\cite{elkhatib2016mapping}, CSPs might use each other's resources to offload peak loads or to move processing closer to the end clients. 
Our principal contribution is an algorithm and a prototype implementation of a distributed broker that fairly balances the load between independent CSPs. 
\blfootnote{Author's version of a paper accepted to Cross-Cloud 2018 (EuroSys Workshop). The final version is available at ACM via \acmurldoi
}

Our notion of fairness stems from the Shapley value, a fairness concept widely used in game theory and economics.
Informally, the Shapley value of an agent is equal to her relative contribution to the common good.
The goal of our system is, essentially, to reward CSPs that provide resources when these resources are truly needed, i.e., when others indeed use them. 
The reward consists of priority treatment of loads of such ``accepting'' CSPs when they, in turn, get overflowed or prefer non-local processing.

However, our method might be also used for balancing the loads and excess capacities directly between the clients of CSPs. Usually, when the load is low, a client downscales its rented resources. However, in some situations downscaling is hard or sub-optimal: 
a client might have a long-term rent agreement with a CSP; or it might prefer to keep renting resources to get a rebate on the rent rate; or it might have local, bare-metal resources. In such situations, with our methods clients could form grassroots load-balancing agreements, trading  their excess capacities (however, in order to simplify the presentation, even in such scenarios we call the parties trading the capacities as CSPs).

Our method is non-monetary, in contrast to monetary spot markets used now to trade excess, short-term CSP capacity by some providers~\cite{kash2016economics}.
A client could use spot instances to dynamically migrate between CSPs~\cite{jia2016smart}. However, spot markets require the CSP to set and dynamically manipulate the price, which is a non-trivial problem. 

Our broker is decentralized. Individual CSPs run their local brokers which communicate with each other through a coordination layer. When a CSP wants to migrate a task, the broker submits the task to the global queue. When another CSP has free resources, its broker uses our CloudShare algorithm to choose tasks from the global queue. The cloud API is not exposed outside the CSP: it is only accessed by the broker (and perhaps other local submission systems: we don't require an exclusive access).

The local broker is implemented as a standalone Java application with an embedded HTTP server. The prototype supports executing jobs using local CPU or Kubernetes but it could be extended to support other APIs (e.g., Slurm) by implementation of another driver.

The paper is organized as follows. We start by discussing related work in Section~\ref{sec:related-work}. We then formalize our resource management model, show how to apply Shapley value to cross-cloud load balancing and propose the load balancing algorithm in Section~\ref{sec:theory}. In Section~\ref{sec:implementation} we describe our prototype implementation. In Section~\ref{sec:simul-exper} we show results of simulation experiments.

\section{Related work}\label{sec:related-work}
FairShare~\cite{fairshare} is arguably the most popular approach to fair scheduling.
Fairness is based on predefined shares assigned to each user (or a group).
Task's priority is proportional to this share and inversely proportional to the actual (consumed) share.

In our previous work~\cite{shapleyScheduling,fairshareNotEnough} we proposed an alternative approach based on Shapley value~\cite{shapley}. We describe it in detail in Section~\ref{ssec:shapleyValue}. In this paper, we adopt DirectContr, a heuristic proposed in~\cite{fairshareNotEnough} to a decentralized environment and to cloud computing scenario. 

The notion of contribution in DirectContr is similar to reputation used in OurGrid~\cite{OurGrid}. However, our method allows sites to choose tasks (rather than being requested to do some). Moreover, our method is based on and tested against a notion of fairness widely accepted in other fields.

We deliberately focus on a single issue: fairness in cross-site scheduling. To construct a viable prototype, we do not address many orthogonal problems. For instance, we use standard container repositories to instantiate a task, while a complete system should use cross-cloud data storage such as~\cite{rafique2017towards}; or even consider VM migration between perhaps binary-incompatible CSPs.
Similarly, we do not consider the problem of different APIs---we assume that each resource is represented by a driver exposing common functionality (which in a complete system requires a cloud orchestration tool~\cite{baur2016experiences}).

\section{Theory: Model and Algorithms}\label{sec:theory}
This section introduces the theoretical motivation, the algorithm and the architecture of CloudShare.

\begin{figure}[tb]
\centering
\includegraphics[width=8cm]{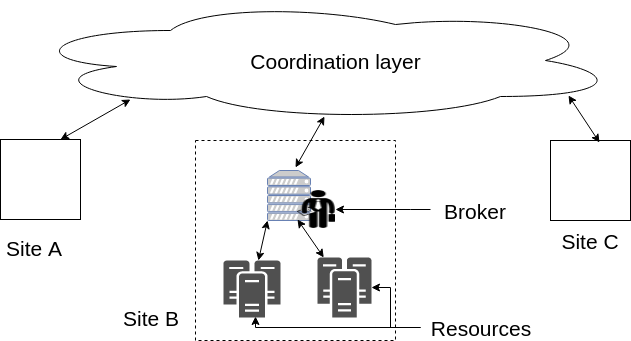}
\caption{ClusterShare overview}
\label{fig:overview}
\end{figure}

\subsection{Vocabulary and Assumptions}
We call a \emph{federation} a system composed of multiple CSPs that balance the load.
We call an individual CSP also an \emph{organization} (a term common in the theoretical works).
Each CSP (organization) has a certain number of \emph{machines} that correspond to physical machines or VMs rented on the long-term.
We assume a machine has a certain number of CPU cores (to simplify our theoretical model, we assume that the CPU is the sole resource
--- however it is easy to generalize our approach to multiple resources).
Each CSP processes \emph{jobs} that are initially submitted locally to this CSP (e.g. by the end clients).
We consider an \emph{on-line} problem with \emph{non-zero} release dates: a job is not known until the moment it is submitted to a CSP.
Each job declares the number of CPU cores it requires exclusively (such declaration is equivalent to, e.g., VM capacity or resource requirements in Kubernetes).
Each job will be executed on a single machine, but a single machine can execute multiple jobs at the same time (with no overbooking of the available cores). 
Jobs have finite duration, but the scheduler does not know the job's duration until the job completes (finishes) (a \emph{non-clairvoyant} problem).
Each organization uses a \emph{utility function} as a performance measure (e.g., the average flow time).




Cooperation of CSPs requires some level of trust. We assume that a CSP does not try to tamper jobs, i.e., all results are genuine outcomes of job execution. In general, verification of a result may require performing the same computation. Therefore it does not make sense to ask untrusted party to run a job. Similarly all metadata (e.g. job start and completion time) must be true. An organization might simulate long execution of a job by delaying the result announcement and supplying false time stamps. This could artificially increase priority of the organization and it would be hard to detect, as it is difficult to predict duration of a job based solely on its definition.
We also assume that CSPs do not alter broker implementations or the data (these problems are orthogonal to the main issue).




\subsection{Fairness based on Shapley Value} \label{ssec:shapleyValue}
Following our earlier theoretical works~\cite{shapleyScheduling,fairshareNotEnough}, we base our notion of fairness on the Shapley value. The main difference from FairShare (see Section~\ref{sec:related-work}) is that share entitlements are not predefined. Instead, they depend on CSP's impact on the federation. The aim is to promote organizations which provide resources when they are needed by assigning a higher priority to their jobs. Calculation of target shares is based on game theoretical concept of Shapley value 


\subsubsection{Shapley value}

A concept from game theory, the Shapley value~\cite{shapley} can be interpreted as a value that a member brings to the community (\emph{a coalition}). The formulation assumes there is a characteristic function $f$ which assigns a value to every subset of possible coalition members, $N$. Shapley value $\phi_o$ of organization $o$ is:

\begin{equation}\label{eq:shapley}
  \phi_o(f)=\sum_{S \subseteq N \setminus \{o\}} \frac{|S|!\; (n-|S|-1)!}{n!}(f(S\cup\{o\})-f(S))\text{.}
\end{equation}

Thus, the Shapley value of $o$ is essentially its average marginal \emph{contribution} to coalition value: the difference between the value of the characteristic function for a subset including the member and the same subset excluding the member. It has desirable properties of efficiency, symmetry, linearity and assigns $0$ to the members who do not contribute anything to the coalition.


As an illustration, for simplicity assume that the characteristic function $f$ is the difference between the number of completed and submitted jobs.
Two events change the Shapley value $\phi_o$: submitting a job or completing it.
An organization which does not submit or complete any tasks will have zero Shapley value. An organization that has completed more tasks than it (locally) submitted will have a positive Shapley value; and an organization only submitting tasks, but not accepting any tasks will have a negative Shapley value.

A scheduling algorithm uses Shapley value as a benchmark. Ideally, the value of the organization's utility function should be equal to its Shapley value, $\phi_o$. However, as the problem is discrete, it might be not possible to achieve such a schedule (e.g.: an organization does not submit any jobs, but accepts jobs from others). Thus, the goal is to construct a schedule with utilities as close to Shapley values as possible (see~\cite{shapleyScheduling} for a more formal discussion).

\subsubsection{Utility/characteristic functions}\label{sec:util-funct}
The Shapley value relies on the utility function $f$ quantifying the quality of the schedule from  the coalition's perspective. 

In our previous work~\cite{shapleyScheduling}, we proposed a non-manipulable utility function that is the sum of utilities for individual jobs~$j$ computed as:
\begin{equation}
  \psi(j) = (e(j) - s(j)) \cdot cpu(j) \cdot \Big(T - \frac{s(j) + e(j) - 1}{2}\Big)\text{,}
\end{equation}
where $s(j)$ denotes the time the job $j$ started; $e(j)$---job $j$ ended, $T$---the current time and $cpu(j)$---the number processor cores $j$ used (an extension we added here, as~\cite{shapleyScheduling} considered only sequential jobs). The utility is proportional to the duration and depends on the start time. 

The term $\Big(T - \frac{s(j) + e(j) - 1}{2}\Big)$ expresses the time since the job is run. The motivation was to capture utility of having the same amount of work done faster. However, in a long running system, it might be impossible to ``make up'' for the sub-optimal decisions taken at the beginning of the schedule. Therefore, we also test a slightly altered utility, where the release time $r(j)$ neutralizes this effect:
\begin{equation}
\psi'(j) = (e(j) - s(j)) * cpu(j) * \Big(T + r(j) - \frac{e(j) + s(j) - 1}{2}\Big)\text{.}
\end{equation}

Finally, we also consider a function that sums the surface of executed jobs (with no reward for executing a job earlier):
\begin{equation}
  \psi''(j) = (e(j) - s(j)) * cpu(j)\text{.}
\end{equation}
While $\psi''$ does not adequately express utility, it is reasonable in expressing contribution---the effort of a site that accepts non-local jobs.


\subsubsection{DirectContr: Scheduling based on Shapley Value}
Calculating the Shapley for an organization is NP-hard and hard to approximate~\cite{shapleyScheduling}.
We proposed a fast heuristic called \emph{DirectContr}~\cite{fairshareNotEnough}.
Instead of computing the Shapley value from the definition (Eq.~\ref{eq:shapley}),
the algorithm estimates the contribution of an organization $o$
by summing utilities from jobs 
executed on $o$'s resources.
The algorithm works as follows.
An organization $o$ submits its jobs to its queue $Q_o$.
Each time a processor becomes available,
the algorithm selects the organization $o^*$ that
has the highest difference between its contribution and its utility---we will call this difference the \emph{priority} (if there are multiple free processors, the algorithm selects one randomly).
Then, it executes the first job from this organization's queue, $Q_{o^*}$.

\begin{figure}[tb]
\centering
\includegraphics[width=8cm]{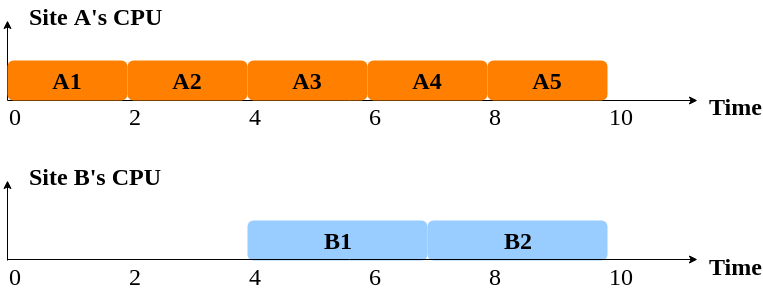}
\caption{DirectContr schedule}
\label{fig:dcs}
\end{figure}

\begin{figure}[tb]
\centering
\vspace{-1em}\includegraphics[width=8cm]{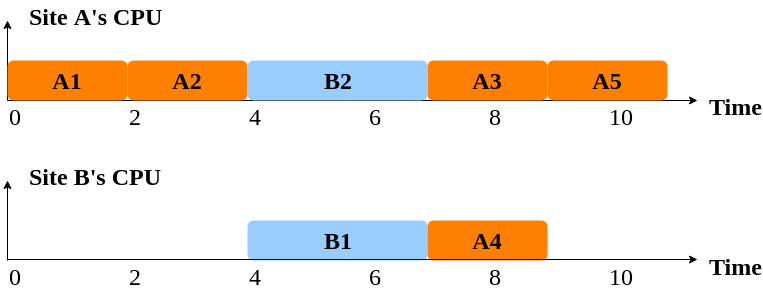}
\caption{FairShare schedule}
\label{fig:fss}
\end{figure}

By simulation, in~\cite{fairshareNotEnough} we showed that the ``unfairness'' of the resulting schedule is relatively close to the exact, exponential algorithm, and significantly lower than the FairShare.
This result can be intuitively explained on an example (see Figures \ref{fig:dcs} and \ref{fig:fss}).
Consider two organizations \emph{A} and \emph{B}, each with a single machine. \emph{A} submits a job at time 0, 2, 4, 6 and 8. \emph{B} submits two jobs at time 4.
Consider the situation at time 4.
In DirectShare (\autoref{fig:dcs}), priorities of \emph{A} and \emph{B} are equal: jobs were executed on local resources, thus both organization have 0 contribution.
In contrast, in FairShare, \emph{B} has higher priority: both organizations have the same predefined share but all completed jobs belong to organization \emph{A}.
FairShare scheduler would decide to start both jobs submitted by \emph{B} immediately (\autoref{fig:fss}). In the aftermath jobs \emph{A3}, \emph{A4} and \emph{A5} are delayed 5 time units compared to the DirectContr schedule.
Moreover, we could multiply the number of processors or extend the job duration to arbitrarily large total delay of organization \emph{A}'s jobs.

\subsection{Distributed Scheduling in ClusterShare}

ClusterShare adapts DirectContr to the federated cloud infrastructure.
An organization (a CSP) is represented in the system by a broker, responsible for tracking local resources, submitting local tasks to the federation, selecting and executing foreign tasks on the local resources.

ClusterShare keeps the state of the system in a \emph{coordination layer}, a distributed data structure shared across brokers. The coordination layer keeps track of non-local jobs' life-cycle and execution parameters. 


In contrast to DirectContr, scheduling decisions in ClusterShare are distributed: each broker reacts to events independently.
This approach has several advantages: (i) the decision to expose local resources might be taken dynamically; (ii) local resource schedulers may pursue custom goals like power efficiency; (iii) resources can be exposed to the federation through existing interfaces.


ClusterShare is event-based. Every broker handles events sequentially in order of their appearance. The following events are handled:
\begin{enumerate}[leftmargin=0.5cm,topsep=0cm]
\item a new task is submitted by a local user;
\item a new task is submitted to the federation;
\item a local resource is ready to execute a task;
\item a local resource completed a task;
\item a site left the federation.
\end{enumerate}

When a task is submitted by a local user (event 1), the broker first checks whether the site has enough resources to run the job locally. If there is enough capacity and no other tasks are waiting then the task is delegated to a free machine directly. Otherwise the broker publishes the task in the coordination layer.

When a new task is published in the coordination layer (event 2), the broker propagates the event to every configured resource handler. Some handlers will wait until there is enough capacity to run the job, others will submit a pilot job in order to acquire them.

Once a resource handler is ready to execute a job (event 3), it notifies the broker, which picks a federation job (we show the algorithm below).
In our implementation, a local machine is exposed to the federation (and thus event 3 is produced) when two conditions are met: (i) the overall reservation ratio is below a certain threshold (e.g.: 30\% of the total CPU power is not reserved by currently executing tasks); (ii) there is at least one machine capable of executing a task. The threshold is used to neutralize uncertainty resulting from monitoring delay. This strategy is especially useful for resources which discard jobs that they cannot accept due to insufficient capacity rather than appending them to a queue (like Kubernetes).
\fixme{milpac: I find it a strong limitation. On the bright side we don't need (i) if we exclude unmanaged local traffic}

Continuing with handling event 3, the broker first loads all waiting tasks from the coordination layer and filters those that match the offer.
Then, a task is chosen according to a strategy based on DirectContr, i.e., first, an organization is picked according to the priority (the difference between the contribution and organization's utility); then, the longest waiting task of this organization is chosen. 
Once the job is selected the broker registers itself as the provider via coordination layer and initiates job execution. 

When a task ends (event 4), a resource handler notifies the broker. The broker updates the job status in the coordination layer and saves scheduling-relevant parameters such as the start time, the end time and the job definition.

When organization looses connection to the federation (event 5), jobs that were computed by the disconnected organization are treated as if they were just submitted to the federation to ensure that they will be rescheduled. Jobs submitted by the lost site are removed, apart from those which are already being executed.

\section{Implementation}\label{sec:implementation}
In this section we describe how we map the architecture and the algorithm described above to a hybrid cloud infrastructure. Our implementation is composed of three logical layers: (i) the coordination layer responsible for storing the data shared across CSPs; (ii) the application layer implementing the ClusterShare algorithm; and (iii) the external layer abstracting the resources our algorithm manages. 










The \emph{coordination layer} keeps track of non-local jobs' life-cycle and execution parameters. To increase resilience, we use Apache ZooKeeper~\cite{ZooKeeper}, a well-known, distributed open-source coordination service. ZooKeeper servers can run on the same machines that organizations use to expose their brokers or they can be deployed on separate machines. Brokers use ZooKeeper clients to communicate with ZooKeeper servers. It is easy to add new CSPs to the hybrid cloud: all a new member has to do is to connect to the already established ZooKeeper ensemble.






The \emph{application layer} is composed of symmetrical brokers---each broker represents a CSP. A broker manages only the resources of its CSP. Brokers use the coordination layer to communicate across CSP boundaries. A broker also exposes an HTTP interface for accepting jobs from its local users. 


Each of CSP resources (e.g., a Kubernetes-managed cluster, or a Slurm-managed cluster) is represented through a resource handler with a common interface. 

To test various scheduling algorithms, we abstract any algorithm through an interface with a single method. 
The method, given a collection of jobs, selects the one with the highest priority (priority calculation depends on the algorithm).


As ZooKeeper, our coordination layer, is not well-suited for storing large data, a scheduling algorithm periodically replaces the historic scheduling data with a summary that allows to calculate priorities in the future without resorting to the original release/completion times. 


The \emph{external layer} includes resources managed by ClusterShare brokers and peripheral services (such as broker's client interface or container libraries).  
To instantiate our system, we focus on sharing Kubernetes-managed clusters. The broker uses Kubernetes to start a task, monitor its progress and also monitor the state of the resources (e.g.: whether there are free resources to start a foreign task). We use one-to-one mapping between ClusterShare tasks and Kubernetes job definition: container image name, command arguments, and resource requirements are copied directly.
ClusterShare does not directly manage the containers, nor the eventual results. We envision a setup where CSPs share access to container image repositories and perhaps file repositories. An image should describe both the actual task and result delivery. \fixme{krz to milosz: 2 sentences here what to do about it; milosz: added, pls see comment below}




\section{Simulation Experiments}\label{sec:simul-exper}
To quantitatively evaluate ClusterShare, we performed simulation experiments in which we compared the performance of a number of algorithms.

\subsection{Method}
As our goal was to evaluate the system in steady-state and on workloads lasting days rather than minutes, instead of emulation, we decided to implement a simple, event-based simulator. Our simulator replaces the original external and coordination layers of ClusterShare with fast, in-memory implementations; the broker module is the same as in the ClusterShare system. The simulator also allows us to submit tasks at each site according to a \emph{workload} (log) and then to compute the performance of the schedule.

To compare how ``good'' a particular schedule is, for each CSP we compute the total wait time of tasks (the time between a task is submitted and it is started).
It is an intuitive measure: improved wait time can be the primary reason to federate in a hybrid cloud.
However, the wait time might be sensitive to even minor changes in the schedule (for instance, a long task scheduled a unit time earlier might delay a large number of other tasks).
Moreover, rather than using the wait time directly, we compute the unfairness, or the distance between the resulting vector of waiting times and the perfectly fair vector (the Shapley value).
This requires that the sum of utilities is constant (i.e.: all possible schedules have the same total wait time; only the distribution of the wait time across CSPs changes), because it is interpreted as a characteristic value of a coalition (explained below) which should be fixed for a given log sample.
For these reasons, we convert tasks in the logs by, first, replacing a $q$-processor task with $q$ tasks requiring a single processor; and then replacing a task lasting $p$ hours with $\lceil p \rceil$ tasks each lasting an hour.
We stress that this processing is done to emphasize the differences between scheduling algorithm policies: this method only reduces the noise in the observed results.

Given a log, for each tested algorithm we run $2^N - 1$ simulations (where $N=5$ is the number of CSPs). Thus, there are 5 simulations in which each CSP uses only its local resources;
\fixme{krz to milosz: nie jest jasne: w nastepnym zdaniu piszemy ze each simulation yields a vector of length N; ale robimy 5 symulacji dla pojedynczych organizacji; czy to nie zbyt duzy skrot myslowy?}
10 simulations for pairs of collaborating CSPs, \dots, and one simulation for the grand coalition of 5 CSPs. Each simulation yields a vector (of length N) of total wait time. The sum of this vector is the characteristic value of the coalition, since we assumed that total wait time is our characteristic function. We use all $2^N - 1$ vectors to compute the Shapley value for each coalition (Eq.~\ref{eq:shapley}). Thus we can evaluate the unfairness of a schedule by comparing total wait times of organizations with their Shapley values.

We aggregate results across logs as follows. We assign a score to each algorithm based on how it performs in comparison to other algorithms. The score is incremented every time an algorithm is more fair than another algorithm. For instance, if we test 3 algorithms A, B and C, and the simulator generated three schedules $\sigma_A$, $\sigma_B$ and $\sigma_C$ with their respective unfairness of 60, 80 and 60, then we assign 1 point for A and 1 point for C (an alternative, the average deviation from the fair wait time distribution, might vary significantly across different logs).


We use HPC2N, DAS2 fs0, LPC EGEE and MetaCentrum logs from~\cite{feitelson2014experience}. From each log, we take 20 randomly-chosen periods of 24 hours; we take all jobs submitted during these 24 hours, keeping their relative release dates. Each job is local to some CSP; as in a log each job has an owner user ID. We map these user IDs onto CSPs (and assign number of processors to each CSP) as follows:
\begin{description}[leftmargin=0cm,topsep=0cm]
\item[Scenario 1] CSPs have equal number of processors; users are randomly assigned to CSPs.
\item[Scenario 2] distribution of processors across CSP follows the Zipf law; users are randomly assigned to CSPs.
\item[Scenario 3] CSPs have equal number of processors; users are divided into two categories: ClusterShare users who submit to the broker (as in previous two scenarios); and users local to each CSPs generating background load.
\end{description}

\subsection{Algorithms}
We compare a few variants of ClusterShare with more classic approaches:
\begin{description}[leftmargin=0cm,topsep=0cm]
\item[Original direct contribution] (\texttt{ORIG\_DIRECT}) implements a distributed version of DirectContr with $\psi$ utility function (Section~\ref{sec:util-funct}). 
\item[Relative direct contribution] (\texttt{REL\_DIRECT}) uses $\psi'$ utility (adjusting utility by release time).
\item[Simplified direct contribution] (\texttt{SIMPL\_DIRECT}) uses $\psi''$ utility; this algorithm tests whether direct contribution algorithm could be simplified without sacrificing fairness.
\item[FairShare] uses shares proportional to the number of processors a CSP contributes.
The algorithm measures the total processing time assigned to each organization (just as  $\psi''$). The CSP with the the smallest ratio of utility to share has the highest priority.
\item[Round robin] implementation is based on LRU cache invalidation. The algorithm selects the CSP which has never been chosen yet or (if all were selected at least once) the one whose most recently started job was started least recently.
\end{description}

\begin{table}[tb]
  \small
    \centering
    \begin{tabular}{ l||r|r|r }
    \textbf{Algorithm} & \textbf{Scenario 1} & \textbf{Scenario 2} & \textbf{Scenario 3} \\
    \hline 
    \texttt{ORIG\_DIRECT} & 3051 & 3095 & 2110 \\
    \texttt{REL\_DIRECT} & 3840 & 3793 & 2420 \\
    \texttt{SIMPL\_DIRECT} & 4061 & 3955 & 2415 \\
    \texttt{FAIRSHARE} & 3108 & 3436 &  1621 \\
    \texttt{ROUND\_ROBIN} & 1302 & 917 & 871 \\
    \end{tabular}
    \caption{Total scores: how many times a given algorithm is more fair than another algorithm}
    \label{fig:allScenarios}\vspace{-3em}
  \end{table}

\subsection{Results}
  We summarize the scores in Table~\ref{fig:allScenarios}.
 First, while \texttt{ROUND\_ROBIN} scored significantly less points then other algorithms, its result is not 0: on some logs and some scenarios, \texttt{ROUND\_ROBIN} performs better than seemingly more fair algorithms.

  Second, DirectContr produces more fair schedules than FairShare. The scores of  \texttt{SIMPL\_DIRECT} and \texttt{REL\_DIRECT} were better than scores of \texttt{FAIRSHARE} in every scenario and every analyzed configuration. \texttt{SIMPL\_DIRECT} version of DirectContr unexpectedly achieved the highest total score in Scenarios 1 and 2. The advantage over \texttt{REL\_DIRECT} is not very significant though.  \texttt{ORIG\_DIRECT} performed well in some configurations but its total score is below the score of FairShare in the first two scenarios.

  Third, when the amount of contributed resources changes dynamically (Scenario 3),
the number of processors does not correspond to the contribution.
As, DirectContr does not assume fixed shares (unlike FairShare),
thus 
the advantage of DirectContr is more visible.

\vspace{-1em}
\section{Conclusion}
ClusterShare is a prototype system for fair resource sharing in a hybrid cloud. Our main objective was to adapt DirectContr algorithm to a distributed system and cloud computing scenario.
Clustershare was designed to be flexible and resilient. There is no need for common federation servers --- deployment on private servers of organizations is possible. The federation can grow or shrink spontaneously without disrupting the system. Each site is responsible for its own resources only but it picks jobs from the common queue based on the global priority of each site. 
Container engine guarantees job portability between sites and resources.

We verified performance of our method by simulation. Our results show that methods based on the Shapley value lead to more fair outcome distribution than FairShare.





\small{\noindent \textbf{Acknowledgements:} This research has been partly supported by
the Polish National Science Center grant Sonata
(UMO-2012/07/ D/ST6/02440), and project TOTAL that has received funding
from the European Research Council (ERC) under the European Union's
Horizon 2020 research and innovation programme (grant agreement No
677651).}

\bibliographystyle{abbrvnat}
\vspace{-1em}
\bibliography{references}

\end{document}